# STOI-Net: A Deep Learning based Non-Intrusive Speech Intelligibility Assessment Model

Ryandhimas E. Zezario*†, Szu-Wei Fu†, Chiou-Shann Fuh*, Yu Tsao†, Hsin-Min Wang‡
*Department of Computer Science and Information Engineering, National Taiwan University, Taipei, Taiwan
E-mail: fuh@csie.ntu.edu.tw
†Research Center for Information Technology Innovation, Academia Sinica, Taipei, Taiwan
E-mail: {ryandhimas, jasonfu, yu.tsao}@citi.sinica.edu.tw
‡Institute of Information Science, Academia Sinica, Taipei, Taiwan
Email: whm@iis.sinica.edu.tw

*Abstract*— The calculation of most objective speech intelligibility assessment metrics requires clean speech as a reference. Such a requirement may limit the applicability of these metrics in real-world scenarios. To overcome this limitation, we propose a deep learning-based non-intrusive speech intelligibility assessment model, namely STOI-Net. The input and output of STOI-Net are speech spectral features and predicted STOI scores, respectively. The model is formed by the combination of a convolutional neural network and bidirectional long short-term memory (CNN-BLSTM) architecture with a multiplicative attention mechanism. Experimental results show that the STOI score estimated by STOI-Net has a good correlation with the actual STOI score when tested with noisy and enhanced speech utterances. The correlation values are 0.97 and 0.83, respectively, for the seen test condition (the test speakers and noise types are involved in the training set) and the unseen test condition (the test speakers and noise types are not involved in the training set). The results confirm the capability of STOI-Net to accurately predict the STOI scores without referring to clean speech.

## I. Introduction

For many speech-related applications, such as assistive oral communication devices [1-5] and telecommunications [6-8], and speech-related tasks, such as speech coding [9, 10], voice conversion [11, 12], speech separation [13, 14], and speech enhancement [15-18], speech intelligibility plays a crucial role in determining the performance of processed speech signals. An intuitive method to measure speech intelligibility is to conduct a human listening test. By playing test samples to subjects, the intelligibility scores can be calculated by the ratio of the number of accurately recognized words to the total number of words in the played speech samples. To make an accurate and unbiased evaluation of speech intelligibility, it is necessary to recruit as many subjects as possible, and each subject must listen to a large number of test utterances covering diverse conditions. In general, this may be prohibitive and may not be feasible. Therefore, several approaches have been proposed to estimate speech intelligibility as surrogates for the human listening test [19-23].

The articulation index (AI) [19] and speech intelligibility index (SII) [20] are two well-known objective speech intelligibility predictors; both metrics have been widely used to measure speech intelligibility in various speech-related applications. Based on the design of SII, extended SII (ESII) [24] and coherence SII (CSII) [25] were derived to attain better intelligibility measurements. The speech transmission index (STI) [21, 22] extends the range of distortion to convolutive noise (e.g., reverberant speech and effects of room acoustics) by considering the depth of temporal signal modulation compared to the clean, undistorted reference signal. Recently, short-time objective intelligibility (STOI) [23] has been proposed. Its calculation consists of five major steps: (1) silent frame removal, (2) short-time Fourier transform (STFT), (3) one-third octave band analysis, (4) normalization and clipping, and (5) intelligibility measurement. In terms of predictive ability, STOI has shown a notable improvement in intelligibility scoring in several domains [26-28] over the previous methods, and has therefore been widely used as a standardized evaluation metric for many speech-related tasks. A notable limitation of STOI, however, is the requirement for clean speech as a reference, which may not always be accessible, especially during online operations. Several extensions to address this issue have been developed, such as non-intrusive STOI [29].

Although the traditional signal processing-based intelligibility assessment metrics have shown satisfactory measurement results and have been widely adopted as assessment tools for various speech-related tasks, their applicability is still limited because of the following two factors. (1) The generalization of these metrics to new conditions still has room for improvement. Particularly, the assessment performance may degrade while operating under new and unseen conditions. Even if some training data for the new conditions are available, the assessment metrics cannot be adapted to the new data. (2) The compatibility of these metrics to speech processing systems, which are usually built based on deep neural networks in recent years, is restricted. More specifically, these traditional metrics cannot be readily integrated with speech-related systems (such as noise reduction and speech separation) to jointly optimize the overall performance. Due to the above limitations, it is crucial to determine an effective intelligibility assessment metric that can be continuously learned to adapt to new test conditions and can be easily combined with (learning-based) speech processing systems.

In our previous work [30], we had developed a neural network-based non-intrusive quality assessment model, namely

Quality-Net, to estimate the perceptual evaluation of speech quality (PESQ) score [31]. Quality-Net has shown remarkable performance in evaluating noisy and processed speech without the need for clean speech as a reference. As Quality-Net is based on deep neural network architecture, its prediction ability for new environments can be improved by adapting to the corresponding new data. Moreover, several studies have combined Quality-Net with speech-related systems to jointly optimize the overall performance [32, 33]. Along with this research direction, this study investigates and develops STOI-Net, a deep neural network model that can accurately predict STOI scores without the need for clean speech as a reference.

The proposed STOI-Net is formed by the combination of a convolutional neural network and bidirectional long short-term memory (CNN-BLSTM) architecture with a multiplicative attention mechanism. The CNN is used to extract informative features from the input data, and the BLSTM is used to model time-variant characteristics. The attention mechanism aims to boost the performance by focusing on important regions while calculating intelligibility scores. Experimental results reveal that the predicted scores yielded by STOI-Net have rather high correlation with the ground-truth STOI scores when tested in both seen (the test speakers and noise types are involved in the training) and unseen (the test speakers and noise types are not involved in the training) conditions. It may be noted that the STOI calculation requires clean speech as a reference, whereas STOI-Net does not. The results confirm the decent ability of STOI-Net to accurately predict STOI scores (evaluation of speech intelligibility) without the need for clean speech as a reference.

The remainder of this paper is organized as follows. Section II introduces the proposed STOI-Net. Section III describes the experimental setup and results. Finally, conclusions and future work are presented in Section IV.

## II. STOI-NET

In this section, we introduce the model architecture and training objective of the proposed STOI-Net.

### A. Architecture

Fig. 1 shows the overall architecture of STOI-Net, which consists of several stages. The input to STOI-Net is a sequence of spectral features of noisy/processed speech, and the output is the predicted STOI score. In STOI-Net, the CNN module has 12 convolutional layers, which are used to obtain informative features from the spectral features. Next, the BLSTM module is used to further model the temporal characteristics of the extracted features from the CNN. The attention mechanism is used to identify and weight the important regions in the input features. In our implementation, multiplicative attention is used to form the attention layer because of its high efficiency and satisfactory performance [34]. Next, a fully connected layer is used to map the frame-wise features into frame-wise scores. Finally, based on these estimated frame-level scores, a global average operation is applied to calculate the final predicted STOI score.

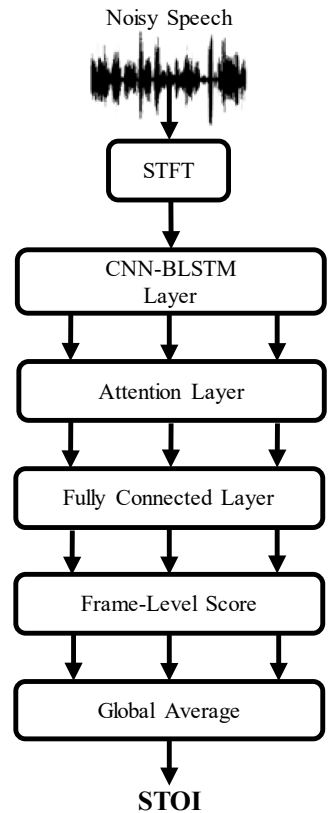

Fig. 1: Architecture of the STOI-Net model.

### B. Objective Function

STOI-Net aims to estimate an utterance-level intelligibility score. However, because a speech utterance may contain non-stationary noises or distortions in different regions (segments of frames), directly assigning an utterance-level score to train STOI-Net may not be a suitable approach. Therefore, we prepare the frame-level scores to train STOI-Net. With the frame-level scores, the objective function can be derived as follows.

$$O = \frac{1}{N}\sum_{n=1}^{N}[(I_n - \hat{I}_n)^2 + \frac{1}{L(U_n)}\sum_{t=1}^{L(U_n)} \alpha(I_n)(I_n - \hat{\imath}_{n,t})^2] \quad (1)$$

where $I_n$ and $\hat{I}_n$ are the true and predicted utterance-level STOI scores, respectively; $N$ denotes the total number of training utterances; $L(u_n)$ denotes the number of frames in the $n$-th utterance; $\hat{\imath}_{n,t}$ is the predicted frame-level STOI score of the $t$-th frame of the $n$-th utterance; and $\alpha(I_n)$ denotes the weighting scale, which is determined by the attention mechanism. It can be seen that the first term estimates the accuracy of utterance-level scoring, and the second term estimates the accuracy of frame-level scoring. We believe that with the objective function in Eq. (1), STOI-Net can be trained to model the STOI metric locally and globally.

## III. EXPERIMENTS

### A. Experimental Setup

The Wall Street Journal (WSJ) dataset [35] was used to prepare the training and test sets in this study. The training set of the WSJ dataset contained 37,416 utterances, while the test set of the WSJ dataset contained 330 utterances, all recorded at a sampling rate of 16 kHz. We polluted the training utterances with 100 types of noise [36], covering both stationary and non-stationary noise types, with 31 different SNR levels ranging from -10 to 20 dB with an interval of 1 dB. A pretrained speech enhancement (SE) model was used to process the noisy utterances to obtain enhanced utterances. The SE model was formed by a BLSTM model with two bidirectional hidden layers, each consisting of 300 neurons. We randomly selected 15,000 noisy utterances, 15,000 enhanced utterances, and 1,500 original clean utterances to form the training set for the proposed STOI-Net model. The 1,500 clean utterances were used to let the STOI-Net learn the highest STOI score (i.e., 1.0).

We prepared two test sets to evaluate the STOI-Net model: the seen and unseen test sets. For the seen test set, we randomly selected 2,350 noisy utterances, 2,350 enhanced utterances, and 300 clean utterances from the remaining utterances in the above large training set. Thus, the seen test set contained a total of 5,000 utterances. It may be noted that the speakers and noise types overlap with those in the training utterances, but the utterances are different. For the unseen test set, we randomly selected 300 utterances from the test set of the WSJ dataset. For this test set, the speakers and noise types were different from those in the training utterances. We used four other noise types (car, pink, street, and babble) and contaminated the speech utterances at 6 SNR levels (-10, -5, 0, 5, 10, and 15 dB). Finally, we randomly selected 2,350 noisy utterances, 2,350 enhanced utterances, together with the 300 clean utterances, to form the unseen test set (a total of 5,000 utterances).

Each utterance in the training and testing sets was converted into a 257-dimensional spectrogram by applying a 512-point STFT with a Hamming window of 32 ms and a hop of 16 ms, which was used as the input for the STOI-Net. Three evaluation metrics, namely mean square error (MSE), linear correlation coefficient (LCC), and Spearman's rank correlation coefficient (SRCC), were used to evaluate the predicted STOI scores.

### B. Effect of Model Architecture

First, we analyzed the prediction capability of STOI-Net with different model architectures. In a previous work [30], BLSTM has shown its advantage in modeling time-variant speech patterns. Therefore, we used the BLSTM model as our baseline system in this study. The proposed STOI-Net was formed by a CNN-BLSTM architecture, which included 12 convolutional layers, each consisting of four channels {16, 32, 64, and 128}, a one-layered BLSTM (with 128 nodes), and a fully connected layer (with 128 neurons).

Table 1 presents the LCC, SRCC, and MSE results of the BLSTM and CNN-BLSTM models under the seen test condition. Higher LCC and SRCC scores denote better results, while lower MSE scores denote better results. From Table 1, it can be seen that CNN-BLSTM outperforms BLSTM consistently, with higher LCC and SRCC scores and a lower MSE score. Table 2 presents the LCC, SRCC, and MSE results of the BLSTM and CNN-BLSTM models under the unseen test condition. In this table, the same trend as in Table 1 can be seen; in other words, compared with the BLSTM baseline, CNN-BLSTM can yield higher LCC and SRCC scores and a lower MSE score. In the following discussion, we will further evaluate the STOI-Net that is developed based on the CNN-BLSTM architecture.

Table 1. LCC, SRCC, and MSE results of BLSTM and CNN-BLSTM under the seen test condition.

|  | LCC | SRCC | MSE |
|---|---|---|---|
| BLSTM [30] | 0.923 | 0.928 | 0.005 |
| CNN-BLSTM | **0.964** | **0.962** | **0.002** |

Table 2. LCC, SRCC, and MSE results of BLSTM and CNN-BLSTM under the unseen test condition.

|  | LCC | SRCC | MSE |
|---|---|---|---|
| BLSTM [30] | 0.764 | 0.784 | 0.029 |
| CNN-BLSTM | **0.789** | **0.797** | **0.016** |

### C. Effect of Attention Mechanism

From the previous experiment, we have confirmed that compared with BLSTM, CNN-BLSTM can achieve better prediction performance for STOI-Net. In this set of experiments, we aimed to further improve the prediction performance by adding a multiplication attention mechanism to STOI-Net; the system is termed CNN-BLSTM$_{ATT}$. From Table 3, it can be seen that under the seen test condition, CNN-BLSTM$_{ATT}$ can achieve higher LCC and SRCC scores compared to CNN-BLSTM. It should also be noted that the MSE score of CNN-BLSTM$_{ATT}$ is lower than that of CNN-BLSTM, but the improvement is small. Table 4 presents the results under the unseen test conditions. The results again show that with the attention mechanism, CNN-BLSTM$_{ATT}$ always produces better LCC, SRCC, and MSE scores, as compared to CNN-BLSTM without the attention mechanism.

Table 3. LCC, SRCC, and MSE results of CNN-BLSTM and CNN-BLSTM-ATT under the seen test condition.

|  | LCC | SRCC | MSE |
|---|---|---|---|
| CNN-BLSTM | 0.964 | 0.962 | 0.002 |
| CNN-BLSTM$_{ATT}$ | **0.970** | **0.968** | **0.001** |

Table 4. LCC, SRCC, and MSE results of CNN-BLSTM and CNN-BLSTM-ATT under the unseen test condition.

|  | LCC | SRCC | MSE |
|---|---|---|---|
| CNN-BLSTM | 0.789 | 0.797 | 0.016 |
| CNN-BLSTM$_{ATT}$ | **0.827** | **0.815** | **0.015** |

To study the reasons for the performance improvement provided by the attention mechanism, we further analyzed the CNN-BLSTM and CNN-BLSTM$_{ATT}$ models by visualizing the hidden layers. As shown in Fig. 2, the representations of the hidden layers of CNN-BLSTM and CNN-BLSTM$_{ATT}$ show different patterns, confirming that the attention layer provides additional weights to specific regions. In addition, the scatter plots of speech intelligibility assessment by STOI-Net are shown in Fig. 3. We compared the scatter plots of the predicted scores generated by CNN-BLSTM and CNN-BLSTM$_{ATT}$. From the figure, it is clear that CNN-BLSTM$_{ATT}$ can predict STOI scores more accurately than CNN-BLSTM. This further shows that the proposed STOI-Net using CNN-BLSTM and the attention mechanism can achieve higher correlation performance, as compared to STOI-Net using CNN-BLSTM without an attention mechanism.

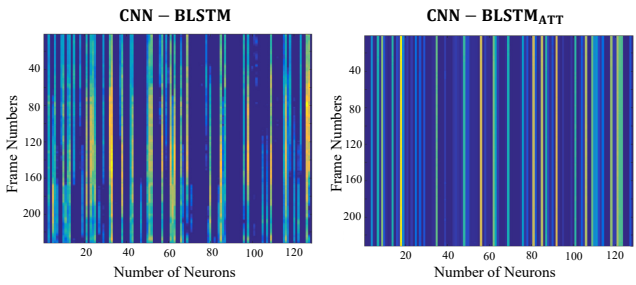

Fig. 2: Representations of a speech utterance at the hidden layers of CNN-BLSTM and CNN-BLSTM$_{ATT}$ models

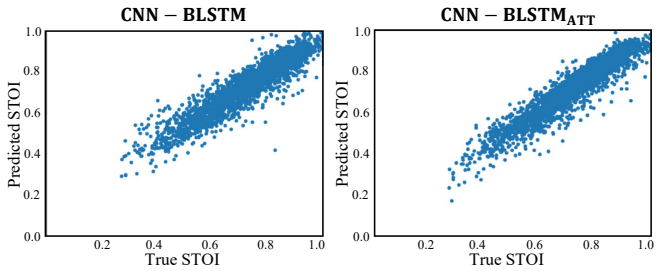

Fig. 3: Scatter plots of speech intelligibility assessment by STOI-Net.

## IV. CONCLUSIONS

In this study, we proposed STOI-Net, a deep neural network-based non-intrusive speech intelligibility assessment model. We aimed to use the STOI-Net as a surrogate to the traditional STOI evaluation metric. Experimental results first confirmed that the predicted scores of STOI-Net have a good correlation with the ground-truth STOI scores. Then, we confirmed the advantages of using CNN-BLSTM over BLSTM to form the STOI-Net model architecture under both seen and unseen test conditions. Finally, we confirmed the effectiveness of the attention mechanism, which can further improve the prediction performance. In the future, we will further evaluate the generalization ability by testing the STOI-Net prediction results under completely different test conditions from those in the training set. We will also explore the integration of STOI-Net into numerous speech-related applications to directly improve the performance of the target tasks.